\documentclass[prb,twocolumn,preprintnumbers,amsmath,amssymb,superscriptaddress]{revtex4}
\usepackage{bm}
\usepackage{amsmath}
\usepackage[dvips]{graphicx}

\begin{document}
\bibliographystyle{apsrev}

\title{Dzyaloshinsky-Moriya-Induced  Order in the
Spin-Liquid Phase of the S=1/2 Pyrochlore Antiferromagnet} 

\author{Valeri N. Kotov} 
\email{valeri.kotov@epfl.ch}
\affiliation{Institute of Theoretical Physics,  Swiss Federal Institute of Technology (EPFL),
1015 Lausanne, Switzerland}
\author{Maged Elhajal}
\affiliation{Max-Planck-Institut f{\"u}r Mikrostrukturphysik,
Weinberg 2, 06120 Halle, Germany}
\author{Michael E. Zhitomirsky}
\affiliation{Commissariat \`a l'Energie Atomique, DSM/DRFMC/SPSMS,
 38054 Grenoble,  France}
\author{Fr\'ed\'eric Mila}
\affiliation{Institute of Theoretical Physics,  Swiss Federal Institute of Technology (EPFL),
1015 Lausanne, Switzerland}

\begin{abstract}
We show that the S=1/2 pyrochlore lattice with both Heisenberg and
antisymmetric, Dzyaloshinsky-Moriya (DM) interactions, can order antiferromagnetically
into a state with chiral symmetry, dictated by the 
distribution of the DM interactions. The chiral antiferromagnetic
 state is characterized by a small staggered
magnetic moment induced by the DM interaction. An external magnetic field can also lead
to  characteristic field-induced ordering patterns, strongly dependent on the
 field direction, and generally separated by a  quantum phase transition from
the chiral ordered phase. The phase diagram at finite temperature is also
 discussed.
\end{abstract}
\pacs{
}
\maketitle

\section{Introduction}
\begin{figure}[ht]
\centering
\includegraphics[height=130pt, keepaspectratio=true]{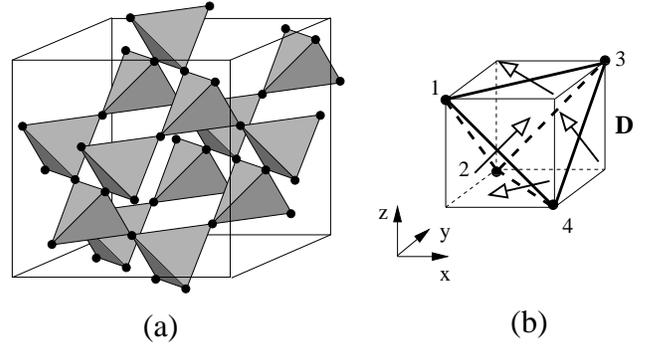}
\caption{(a) Pyrochlore lattice. (b) Distribution of DM vectors on a single
 tetrahedron (four of the six shown, see text).}
\label{Fig1}
\end{figure}

The behavior of many-body systems involving quantum spins
has been one of the central topics in recent years since the
properties of such systems are relevant to a great variety
of materials, mostly oxides. The structure of the ground state
 and the various symmetry broken phases that emerge are issues
 of special interest, especially in systems of low-dimensionality
and/or  where frustration is present. \cite{ML} 
In this context the Heisenberg model on  
the three-dimensional pyrochlore lattice consisting of corner
 sharing tetrahedra, shown  in Fig.~\ref{Fig1}(a), is in a league of its own.
The pyrochlore lattice is strongly geometrically frustrated and is
 relevant to numerous compounds.   
It has been argued that
 no magnetic order is present in the ground state. \cite{HBB,MCCC,T}
The effects of various additional interactions have also been
 studied, such as magnetoelastic couplings, \cite{TMS} 
 long-range dipolar interactions, \cite{PC} and orbital degeneracy. 
\cite{TM} These interactions (in addition to various
 anisotropies) can  generally lead to
bond, magnetic and/or orbital order, and which of them is dominant depends on
the details of the model relevant to the specific class of materials.

 In the present work we  study a new mechanism for magnetic order in
the S=1/2 pyrochlore lattice, driven by the Dzyaloshinsky-Moriya (DM) interactions.
In the pyrochlore such interactions are  expected  to be present by symmetry. 
 For the S=1/2 Heisenberg model on the pyrochlore lattice  it has been suggested \cite{HBB,T}
that  the ground state is dimerized (non-magnetic), but  macroscopic degeneracy
still remains. For certain other lattices, such as the 2D pyrochlore and
related models, \cite{K}  the ground state is a unique valence bond solid,
 and while the DM interactions (if present)  can lead to
 non-trivial order in the ground state, such DM induced order can only occur
 above a critical threshold, due to its inherent competition with the
 underlying dimer order. \cite{K} 
 In this work we show that in the  3D pyrochlore antiferromagnet,
 where a macroscopic degeneracy is present, the DM interactions
 have a more profound effect and can  lift the degeneracy,
 leading to a chiral antiferromagnetic state with a small staggered magnetic moment. 
 In an external magnetic field quantum transitions between weakly ordered states
with different symmetries, depending on the field direction,  
are possible. We determine the field-induced patterns for several 
field orientations, generally pointing in highly-symmetric crystal
directions. The  phase diagram at finite temperature is also
briefly discussed.  
\noindent

\begin{figure}[ht]
\centering
\includegraphics[height=136pt, keepaspectratio=true]{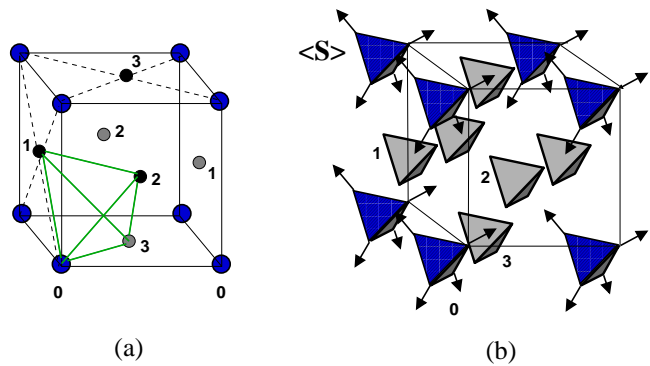}
\caption{(Color online.) (a) Fcc lattice of tetrahedra (tetrahedron = dot) with interactions $J'$ between
them.
(b) Antiferromagnetic chiral order on the blue (dark gray) tetrahedra, with magnetic moment
$|\langle {\bf S} \rangle| \sim \tilde{D}$, induced by the DM interactions. 
On the gray  tetrahedra (labeled as 1,2,3)  the order has the same symmetry,
 but is much weaker $|\langle {\bf S} \rangle| \sim \tilde{D}^{3} \ll  \tilde{D}$,
 Eq.~(9), and is not shown.}
\label{Fig2}
\end{figure}

The spin Hamiltonian (S=1/2) is
\begin{equation}
\hat{{\cal H}} = \sum_{{\bf i,j}}J_{{\bf i,j}} {\bf S}_{\bf i}.{\bf S}_{\bf j}
 + \sum_{{\bf i,j}} {\bf D}_{{\bf i,j}}.({\bf S}_{\bf i} \times {\bf S}_{\bf j}),
\label{ham}
\end{equation}
\noindent
where ${\bf D}_{{\bf i,j}}$ are the DM vectors, to be specified later.
 We start by summarizing the results for ${\bf D}_{{\bf i,j}}=0$,
 i.e. the Heisenberg case.
Our starting point is the strong-coupling approach, similar to that of Refs.~\onlinecite{HBB,T},
with the lattice divided into
two interpenetrating sub-lattices, one of them formed by
 ``strong" tetrahedra (with exchange $J$),
 connected by ``weak" tetrahedra (exchange $J'$). The ``strong"  tetrahedra then
 form a fcc lattice, as shown in Fig.~\ref{Fig2}(a), where every site  represents
 a tetrahedron, and one can attempt to analyze the structure of the ground state
 starting from the limit $J' \ll J$. 

For $J'=0$ the  tetrahedra are disconnected, and on a single
 tetrahedron the ground state is a singlet and is twofold degenerate.
We choose the two ground states as: 
$|s_1\rangle=\frac{1}{\sqrt{3}}\{[1,2][3,4]+[2,3][4,1]\}$,
$|s_2\rangle=\{[1,2][3,4]-[2,3][4,1]\}$,
where $[k,l]$ denotes
a singlet formed by  the nearest-neighbor spins $k$ and $l$, labeled as in Fig.~\ref{Fig1}(b).
In the pseudo-spin $T=1/2$ representation, so that
$T_{z} = 1/2$  corresponds to  $|s_1\rangle$
and  $T_{z} = -1/2$ corresponds to  $|s_2\rangle$, one finds that
third order is the lowest one contributing to the effective 
inter-tetrahedron Hamiltonian
in the singlet sub-space: \cite{Remark}
\begin{equation}
\hat{{\cal H}}_{\mbox{eff}} = \frac{J'^{3}}{8 J^{2}} \left [\hat{{\cal H}}_{\mbox{eff}}^{(2)}
+ \hat{{\cal H}}_{\mbox{eff}}^{(3)} \right ] + \mbox{Const.},
\end{equation}
where
\begin{equation}
\hat{{\cal H}}_{\mbox{eff}}^{(2)} \! = \!
 \sum_{\langle {\bf i,j}\rangle} 
\Bigl \{ \Omega_{{\bf ij}}^{x} T_{{\bf i}}^{x}T_{{\bf j}}^{x} +
\Omega_{{\bf ij}}^{z} T_{{\bf i}}^{z} T_{{\bf j}}^{z}
 + \Omega_{{\bf ij}}^{xz}
 (T_{{\bf i}}^{x}T_{{\bf j}}^{z} \!+ \!T_{{\bf i}}^{z}T_{{\bf j}}^{x}) \Bigr \} ,
\label{pham2}
\end{equation}

\begin{equation}
\hat{{\cal H}}_{\mbox{eff}}^{(3)} \! = \!
 \sum_{({\bf i,j,k})} \!
\left \{ \frac{1}{3} T_{{\bf i}}^{z} T_{{\bf j}}^{z} T_{{\bf k}}^{z}
\!- \!T_{{\bf i}}^{z} T_{{\bf j}}^{x} T_{{\bf k}}^{x}
\!+ \! \frac{T_{{\bf i}}^{z}}{\sqrt{3}}(T_{{\bf j}}^{x} T_{{\bf k}}^{z}
\!- \!T_{{\bf j}}^{z} T_{{\bf k}}^{x})
\right \}
\label{pham3}
\end{equation}
\noindent
In the two-body part we have defined 
\begin{eqnarray}
&& \Omega_{03}^{x}=\Omega_{12}^{x}=1/2, \Omega_{03}^{z}=\Omega_{12}^{z}=-1/6,  \nonumber \\
&& \Omega_{23}^{z}=\Omega_{01}^{z}=\Omega_{02}^{z}=\Omega_{13}^{z}=1/3,  \nonumber \\ 
&& \Omega_{23}^{xz}=\Omega_{01}^{xz}=-\Omega_{02}^{xz}=-\Omega_{13}^{xz}=1/(2\sqrt{3}). 
\end{eqnarray}
All  remaining $\Omega_{{\bf ij}}=0, \ {\bf i}<{\bf j}$. The site indexes 
${\bf i}, {\bf j}$ refer to the fcc lattice
made of individual tetrahedra, Fig.~\ref{Fig2}(a), and it is sufficient to
 know the interactions on one "supertetrahedron", shown in green
(containing the sites 0,1,2,3). In the  three-body
interaction the indexes run over the values:
 $({\bf i,j,k})=\{(3,2,1),(1,0,3),(2,3,0),(0,1,2)\}$.

On a  mean-field level the  ground state of $\hat{{\cal H}}_{\mbox{eff}}$
is defined by the following averages:
\begin{eqnarray}
&\langle T_{1}^{x} \rangle  =  -\sqrt{3}/4, &\langle T_{1}^{z} \rangle  = 1/4, \nonumber \\
&\langle T_{2}^{x} \rangle =  \sqrt{3}/4,& \langle T_{2}^{z} \rangle = 1/4, \nonumber \\
&\langle T_{3}^{x} \rangle =  0,&  \langle T_{3}^{z} \rangle  =  -1/2, \nonumber \\
&\langle T_{0}^{x} \rangle  = 0, &   \langle T_{0}^{z} \rangle  =   0.
\end{eqnarray}
This means that  while a dimerization pattern sets in on sites 1,2,3,
the pseudospins on the ``$0$'' sites,
 shown in blue in  Fig.~\ref{Fig2}(a) remain ``free'', i.e.
there is no fixed dimer pattern on those sites  and consequently
 a macroscopic degeneracy remains.  \cite{T}
 
One should certainly keep in mind that the  strong-coupling approach
 breaks artificially the lattice symmetry and while one hopes that
 the structure of the ground state is correct even in the isotropic
limit $J'=J$, it is very difficult to assess this by other means
(e.g. exact diagonalizations) at the present time. 
Nevertheless this approach is expected to provide reliable
description of  the ground state properties  as long as the relevant physics
remains in the singlet subspace, i.e. the triplet modes stay high in energy
and no magnetic order is generated, as might be the 
case for the  pyrochlore antiferromagnet due to the strong frustration.
Fluctuations around the mean-field solution, Eq.~(6),  can lift the degeneracy,
leading to unique dimer order. 
However the corresponding degeneracy lifting energy scale is very small, \cite{T} of the order of
 $10^{-3} \beta, \ \beta \equiv J'^{3}/(48 J^{2})$.
A unique (singlet) ground state is also produced if one starts the expansion 
 from a larger cluster of 16 sites, with
 an ordering energy scale (energy gain) of $10^{-2} J$,  extrapolated to the limit
where all couplings are equal. \cite{BAA}

 In what follows we will take the mean-field
 solution as a starting point and  discuss
a physical mechanism, based on the presence of interactions beyond Heisenberg exchange,
 that can lead to the lifting of degeneracy and consequently to (magnetic) order in the ground state.  

\noindent

\section{Chiral antiferromagnetic order induced by the Dzyaloshinsky-Moriya interactions}

Now we consider the effect of the DM interactions \cite{D,M} on the ground state properties.
On a single tetrahedron the DM vectors are distributed as shown in  Fig.~\ref{Fig1}(b),
or explicitly: 
${\bf D}_{13}\! =\! \frac{D}{\sqrt{2}}(-1,1,0)$, ${\bf D}_{24}\! =\! \frac{D}{\sqrt{2}}(-1,-1,0)$,
${\bf D}_{43}\! =\! \frac{D}{\sqrt{2}}(0,-1,1)$, ${\bf D}_{12}\! =\! \frac{D}{\sqrt{2}}(0,-1,-1)$,
${\bf D}_{14}\! =\! \frac{D}{\sqrt{2}}(1,0,1)$, ${\bf D}_{23}\! =\! \frac{D}{\sqrt{2}}(1,0,-1)$.
Here $D$ is the magnitude of the (all equal) DM vectors. The directions of the DM vectors
respect the pyrochlore lattice symmetry and thus the DM interactions are
expected to be always present in the system. \cite{MCL}  Since ${\bf D}_{ij}$ originate from the spin-orbit
 coupling, \cite{D,M} we have  $D \ll J, J'$, and typically the values of
 the DM interactions are several percent of the Heisenberg couplings. 
There are two DM distribution patterns that are equally acceptable on symmetry
 grounds - the one shown in Fig.~\ref{Fig1}(b),  and one with all directions
of the vectors ${\bf D}_{ij}$ reversed (${\bf D}_{ij} \rightarrow -{\bf D}_{ij}$).
 These two cases were named, respectively, 
 "indirect" and "direct" in Ref.~\onlinecite{MCL}. The reader is referred to
that paper for more details on Moriya's rules as applied to the pyrochlore lattice.  
In the extreme quantum case of S=1/2, and within our approach, 
 we have found  that the two allowed (by symmetry) DM distributions
 lead to qualitatively the same physics  (see discussion following Eq.~(8)).
 
Following the strong-coupling approach outlined above for the purely Heisenberg case,
 we have to determine how the singlet
 ground states $|s_1\rangle,|s_2\rangle$  on a single tetrahedron are modified
 by the presence of $D$. Since the DM interactions break the spin rotational invariance,
 they admix triplets to the two ground states, not lifting their degeneracy. \cite{K}
We will also be interested in effects in the presence of an external magnetic field,
 and in this case the field (in combination with the DM interactions) also 
mixes certain triplet states with  $|s_1\rangle,|s_2\rangle$. 
In order to determine the additional contributions to $\hat{{\cal H}}_{\mbox{eff}}$,
 it is convenient to express the spin operators on a singlet tetrahedron, labeled as
 in Fig.~\ref{Fig1}(b), in terms of the pseudospin operators.
For magnetic field
${\bf H}=\frac{H}{\sqrt{2}}(1,1,0)$ (along the $1-3$ bond), assuming
$D \ll J$ and $H \ll J$, we obtain (defining the rescaled
 quantities $\tilde{D}, \tilde{H}$ along the way):  
\begin{eqnarray}
{\bf H}&=&\frac{H}{\sqrt{2}}(1,1,0); \ \ \tilde{D} \equiv D/J, \ \ \tilde{H} \equiv H/J, \nonumber \\
S_{1,3}^x  &= & \mp\frac{2\tilde{D}}{\sqrt{6}}T^{y}
-\frac{\tilde{D}\tilde{H}}{\sqrt{3}}T^{x},
 \  S_{2,4}^x=\mp\frac{2\tilde{D}}{\sqrt{6}}T^{y}+\frac{\tilde{D}\tilde{H}}{\sqrt{3}}T^{x}  \nonumber \\
S_{1,3}^y & = & \mp\frac{2\tilde{D}}{\sqrt{6}}T^{y}
+\frac{\tilde{D}\tilde{H}}{\sqrt{3}}T^{x}, \ S_{2,4}^y=\pm\frac{2\tilde{D}}{\sqrt{6}}T^{y}
 -\frac{\tilde{D}\tilde{H}}{\sqrt{3}}T^{x}  \nonumber \\
S_{1,3}^z & = & \frac{2\tilde{D}}{\sqrt{6}}T^{y} \mp \tilde{D}\tilde{H} T^{z}, \
S_{2,4}^z= -\frac{2\tilde{D}}{\sqrt{6}}T^{y} \pm \frac{\tilde{D}\tilde{H}}{\sqrt{3}}T^{x}
\label{operatorsDM}
\end{eqnarray}
\noindent
The notation  ${\bf S}_{i,j}$  simply combines  in one line the formulas for
 both  ${\bf S}_{i}$ and  ${\bf S}_{j}$, where $i$ and $j$ label sites on
 a tetrahedron (as defined in Fig.~\ref{Fig1}(b)),  while  
the left index in  ${\bf S}_{i,j}$ corresponds to the upper sign on
 the right hand side, and the right index - to the lower sign.
The formulas Eq.~(7) are obtained by using the ground state wave-functions,
written explicitly in Ref.~\onlinecite{K} (Eqs.~(2,5)), to lowest order in $\tilde{D}$ and 
$\tilde{D}\tilde{H}$. For magnetic field in the $z$ direction, the corresponding
 expressions are given  in Appendix A.

First we analyze the case of zero magnetic field (${\bf H}=0$).
Taking into account the connections
between the tetrahedra (green bonds in Fig.~\ref{Fig2}(a)), and
using Eq.~(\ref{operatorsDM}),  we obtain an  additional interaction
 term, so that the full effective Hamiltonian $\hat{{\cal H}}_{\mbox{eff}}^{(DM)}$
 becomes 
\begin{equation}
\hat{{\cal H}}_{\mbox{eff}}^{(DM)}= \hat{{\cal H}}_{\mbox{eff}} - J' \tilde{D}^{2} \frac{2}{3} 
\sum_{{\bf i}<{\bf j}}  T_{{\bf i}}^{y}T_{{\bf j}}^{y},  
\label{phamDM}
\end{equation}
where $\hat{{\cal H}}_{\mbox{eff}}$ is the part originating from
the Heisenberg exchanges, Eq.~(2). 
The above result is  obtained in  lowest, first order in $J'$.
 While extra terms of the same power $J'(D'/J)^{2}$ also
 arise from the DM interactions $D'$ on the inter-tetrahedral bonds,
we find that they only  give a small renormalization of the
energy scale $J'^{3}/(8J^{2})$ in  Eq.~(2) and are, therefore, neglected.

 We would like to also point out that in general the coupling constant in 
$\hat{{\cal H}}_{\mbox{eff}}^{(DM)}$
is not symmetric under $D \rightarrow -D$. To verify this requires a calculation
 of the next to leading order in $\tilde{D}$ in  Eq.~(7). We have found that,
 in the case of zero field $H=0$, the next order present is $\tilde{D}^{2}$, and
one has to substitute in all formulas
 $\tilde{D} \rightarrow \tilde{D} - \frac{3}{4\sqrt{2}} \tilde{D}^{2}$.
 Consequently the same substitution has to be made in the coefficient $- J' \tilde{D}^{2} \frac{2}{3}$
in Eq.~(\ref{phamDM}). Most importantly however  the $T_{{\bf i}}^{y}T_{{\bf j}}^{y}$ structure
 of the interaction is not affected by increasing the strength of $\tilde{D} \ll 1$,
 and from now on we will work with the leading order in $\tilde{D}$.
Therefore the physics (ground state structure) associated with the two DM distribution patterns will be
  the same.  This conclusion might be connected with the fact that
 we have kept only the lowest non-trivial order in the coupling $J'$ in the effective
 Hamiltonian - we have used this as our guiding principle as the difficulties
  associated with the derivation and analysis of  higher orders seem insurmountable.  

We have performed  mean-field calculations of
the Hamiltonian defined by Eqs.~(2,\ref{pham2},\ref{pham3},\ref{phamDM})
 in the unit cell of Fig.~\ref{Fig2}(a), as represented by
 the four sites connected by green lines.  The results can be particularly
simply summarized in the limit $\tilde{D} \ll 1$, which is also the case of physical relevance.
 It is physically clear  that ferromagnetic order in the $T_{{\bf i}}^{y}$ component
is generated on the $``0"$ sites, since no order in the $T_{{\bf i}}^{x,z}$ components
(dimer order) was present on those sites 
without DM interactions (on mean-field level), Eq.~(6).  
Indeed we find $\langle T_{0}^{y} \rangle =1/2$, while for 
  the other sites we have, to lowest non-trivial order in $D$,  
  $\langle T_{{\bf i}}^{y} \rangle \approx 1.8 (D/J')^{2}, {\bf i}=1,2,3$.
From Eq.~(\ref{operatorsDM}) it is then clear that a non-zero average
 of the operator $T_{{\bf i}}^{y}$ corresponds to a finite moment in the ground state,
 with magnitude $|\langle {\bf S} \rangle_{{\bf i}}|=\tilde{D} \sqrt{2} \langle T_{{\bf i}}^{y} \rangle$.
 To summarize:  
\begin{eqnarray}
\langle T_{0}^{y} \rangle & = &1/2, \ \  
\langle T_{{\bf i}}^{y} \rangle \approx 1.8 \frac{D^{2}}{J'^{2}}, {\bf i}=1,2,3
\ \ \ 
\Rightarrow \nonumber\\
|\langle {\bf S} \rangle_{{\bf i}}| & = & \frac{\tilde{D}}{\sqrt{2}},  \mbox{{\bf i}}=0;  \ 
|\langle {\bf S} \rangle_{{\bf i}}| \! \approx \! \frac{3.6}{\sqrt{2}} \tilde{D}\frac{D^{2}}{J'^{2}},
 \mbox{{\bf i}}=1,2,3.
\label{moment}
\end{eqnarray} 
 Here $|\langle {\bf S} \rangle_{{\bf i}}|$ stands for the magnitude of the
 moment on each site of pyrochlore lattice, belonging
to a tetrahedron labeled by the index ${\bf i}$.
From (\ref{operatorsDM}) it follows that the moments point out of the
cube's center (the cube is defined in Fig.~\ref{Fig1}(b)), leading 
to formation of sublattices and the order shown in
Fig.~\ref{Fig2}(b). Since from Eq.~(\ref{moment}) $|\langle {\bf S} \rangle_{{\bf i}}|
/|\langle {\bf S} \rangle_{0}| \sim (D/J')^{2} \ll 1, \ i=1,2,3$, we have neglected the magnetic
order on those tetrahedra.
 
 The antiferromagnetic order of Fig.~\ref{Fig2}(b) corresponds to non-zero scalar chirality
$\chi = \langle {\bf S}_m\cdot({\bf S}_n \times {\bf S}_l) \rangle \neq 0$,
 where $m,n,l$ are any three spins on a given tetrahedron.
 The Ising symmetry $T_{{\bf i}}^{y} \rightarrow -T_{{\bf i}}^{y}$ is 
spontaneously broken in the ground
 state, which in terms of real spins corresponds to the time-reversal symmetry broken
state of Fig.~\ref{Fig2}(b). In this state the two ground state wave functions
$|\Phi \rangle$ and $|\Psi \rangle$ 
(see (A1)) form  linear combinations in the
 "chiral" sector:  $\alpha|\Phi \rangle + i \beta |\Psi \rangle$, 
where $\alpha, \beta$ are real
coefficients ($\alpha^{2}+\beta^{2}=1$).
 This combination is ferromagnetically repeated on every
$T^{y}$ ordered tetrahedron. 
A straightforward calculation shows that both 
$\langle T_{{\bf i}}^{y} \rangle \propto  \alpha \beta, \  \chi \propto  \alpha \beta$. 
The energy gain (per site of  Fig.~\ref{Fig2}(a)) from the formation of the ordered state is
$\Delta E= \langle \hat{{\cal H}}_{\mbox{eff}}^{(DM)} \rangle
- \langle \hat{{\cal H}}_{\mbox{eff}} \rangle  \approx -1.8 J' \tilde{D}^{2}
(D/J')^{2}$.
  The order we have just discussed is in competition with other mechanisms for lifting of the
 degeneracy that could originate from the Heisenberg interactions
 themselves (e.g. fluctuations beyond the mean-field), typically also leading to
 very small energy scales. \cite{T}

\section{Magnetic order induced by external magnetic fields in the
presence of DM interactions}
In the presence of an external magnetic field other possibilities for lifting of
 the degeneracy exist. We will consider three  symmetric field orientations, for which
 the results are particularly transparent. The magnetic field generally leads to  splitting
 of the ground states, which in the language of the pseudospins
 produces an on-site ``effective magnetic field"  $h$ in
 the pseudospin $z$ direction. The effective  Hamiltonian has the form
\begin{equation}
 \hat{{\cal H}}_{\mbox{eff}}^{(H)} = \hat{{\cal H}}_{\mbox{eff}}^{(DM)}
+ h \sum_{{\bf i}}T_{{\bf i}}^{z} + \delta \hat{{\cal H}}_{\mbox{eff}}^{(H)} .
\label{phamfield}
\end{equation}
We consider fields in the $(1,1,0)$ and $(0,0,1)$ directions,
 as well as comment on the case $(1,1,1)$, where 
the axes are defined in Fig.~\ref{Fig1}(b). Using  the wave-functions
in a field we obtain (see Eq.~(A3)):
\begin{equation}
h = \left \{ 
\begin{array}{lll}
\frac{1}{2}D^2H^2/J^{3}, &{\bf H}=\frac{H}{\sqrt{2}}(1,1,0)\\
 - D^2H^2/J^{3}, & {\bf H}=H(0,0,1)\\
 0 , & {\bf H}=\frac{H}{\sqrt{3}}(1,1,1)
\end{array}
\right.
\label{pfield}
\end{equation}
$\delta \hat{{\cal H}}_{\mbox{eff}}^{(H)}$ in (\ref{phamfield}) represents lattice contributions, 
originating from the various combinations in Eq.~(\ref{operatorsDM})
once the tetrahedra are coupled, and also producing terms of   order $D^2H^2$. 
These terms are cumbersome and are not  explicitly written, but their effect
 is taken into account in the (numerical) mean-field implementation within 
 the unit cell  of Fig.~\ref{Fig2}(a). A further discussion appears in Appendix B.

\begin{figure}[ht]
\centering
\includegraphics[height=215pt, keepaspectratio=true]{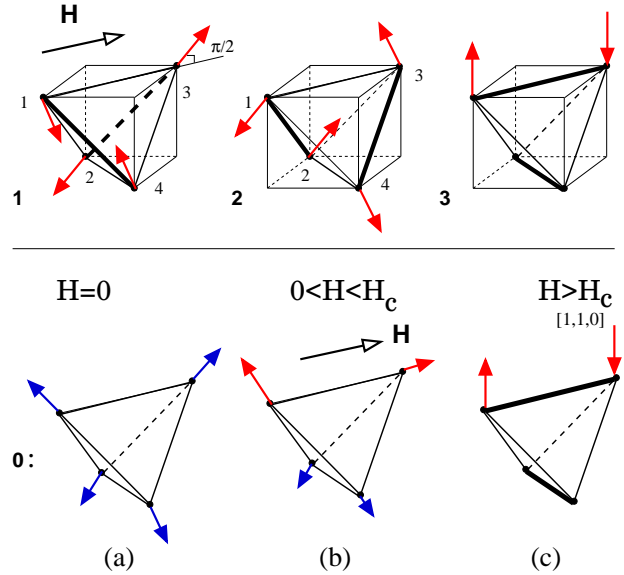}
\caption{(Color online.) (a,b,c) Evolution of magnetic order
on the blue (dark gray)  tetrahedra of Fig.~2(b) in an external magnetic field
 in the $(1,1,0)$ direction. Upper row: field-induced order on the rest
of the tetrahedra. The tetrahedra are labeled 0,1,2,3 as
in Fig.~2(a,b). Blue (dark gray) arrows in (a,b) correspond to moments
 $|\langle {\bf S} \rangle| \sim \tilde{D}$,
while the red (light gray)  arrows on the upper row and (b,c) are the field-induced moments 
$|\langle {\bf S} \rangle| \sim \tilde{D}\tilde{H}$. }
\label{Fig3}
\end{figure}
\noindent

We will mostly discuss the two cases with $h \neq 0$.
Then the on-site $h$ term in Eq.~(\ref{phamfield})  is responsible for the main effect,
namely  competition between  order in the $T_{{\bf i}}^{z}$ pseudospin
 component and  order in the ``chiral" $T_{{\bf i}}^{y}$ component
 favored by Eq.~(\ref{phamDM}). Therefore the physics is that of the transverse field
 Ising model (although in our case the unit cell is larger). It is also
 clear that the mentioned competition is most effective on the
$``0"$ (blue) sites, while the non-zero averages of $T_{{\bf i}}^{z,x}$ on
the other sites are not much affected by the presence of small $D$ and $H$.  
We have found that a quantum transition takes place between a state
 with $\langle T_{0}^{y} \rangle \neq 0, H < H_{c}$ and 
$\langle T_{0}^{y} \rangle = 0, H \geq H_{c}$. The result for
 $\tilde{D}\ll 1$ can be written in an explicit way, and we have 
 for the field ${\bf H}=\frac{H}{\sqrt{2}}(1,1,0)$
\begin{equation}
\langle T_{0}^{y} \rangle^{2} = \frac{1}{4}\left[1-\left(\frac{\tilde{H}}{\tilde{H}_{c}}\right)^{4}\right], 
\ \  \tilde{H} \leq \tilde{H}_{c} \approx 5.3 \sqrt{\frac{J}{J'}} \tilde{D}
\label{critical}
\end{equation}
\begin{equation}
\langle T_{0}^{z} \rangle^{2} = 1/4 - \langle T_{0}^{y} \rangle^{2}
\end{equation}
(and $\langle T_{0}^{z} \rangle <0$ since $h>0$).
The values of the spin moments for given values of $\langle T_{{\bf i}}^{x,y,z} \rangle$
on a tetrahedron can be determined directly from Eq.~(\ref{operatorsDM}).
On the $``0"$ (blue) sites this leads to evolution of the magnetic
order as shown in  Fig.~\ref{Fig3}(a,b,c). 
For  $H=0$ there is only  chiral order (blue arrows)
with moment $|\langle {\bf S} \rangle| \sim \tilde{D}$, changing, for $H>0$ into a combination
of chiral and field induced order (red arrows) with $|\langle {\bf S} \rangle| \sim \tilde{D}\tilde{H}$.
 Gradually, as $H$ approaches $H_c$ the chiral order diminishes (Eq.~(\ref{critical})),
leaving for $H>H_c$ only the field-induced component, equal to
 $|\langle {\bf S} \rangle| = \tilde{D}\tilde{H} |\langle T_{0}^{z} \rangle| = \tilde{D}\tilde{H}/2, \ H>H_c$.   

 On the tetrahedra $1,2,3$ labeled as in  Fig.~\ref{Fig2}(a,b) there is virtually no evolution
as a function of the field, and the order is determined by Eq.~(\ref{operatorsDM})
with $\langle {T_{\bf i}}^{x,z} \rangle$ fixed by the Heisenberg exchanges, see
 Eq.~(6). This leads to the magnetic moments (proportional to $\tilde{D}\tilde{H}$)
 shown  in Fig.~\ref{Fig3}, upper row. On tetrahedra $1$ and $2$ the spins 
point along the internal diagonals of the cube  perpendicular to the field.
Dimerization is also present in the ground state (bolder lines = stronger bonds)
 and co-exists with the magnetic order. 
\begin{figure}[ht]
\centering
\includegraphics[height=210pt, keepaspectratio=true]{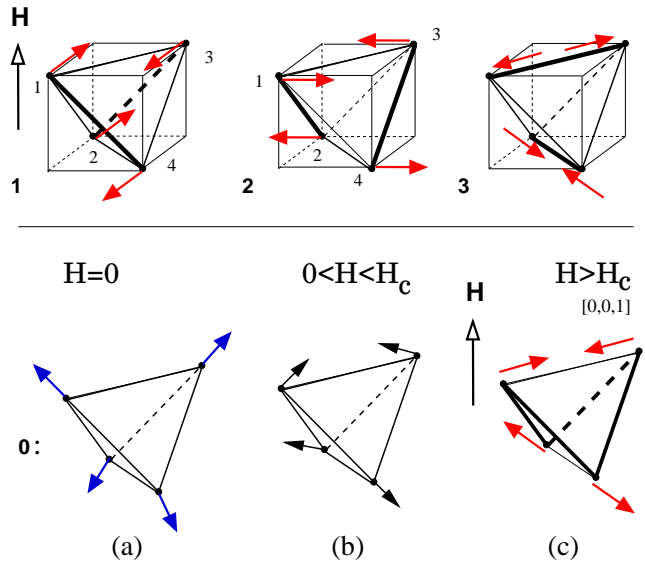}
\caption{(Color online.) Same as Fig.~3, but for 
 an external magnetic field in the $(0,0,1)$ direction.
 Blue (dark gray) arrows in (a) correspond to the
 DM induced order with $|\langle {\bf S} \rangle| \sim \tilde{D}$,
 red (light gray) arrows on the upper row and (c) correspond to the 
 field-induced component $|\langle {\bf S} \rangle| \sim \tilde{D}\tilde{H}$,
and black arrows in (b) are a mixture of the two. }
\label{Fig4}
\end{figure}
\noindent

A similar quantum transition
 takes place  for magnetic field in the $z$ direction ${\bf H}=H(0,0,1)$.
In this case the formulas (A2) from Appendix A have to be used,
 and the field-induced order is shown in  Fig.~\ref{Fig4}. 
 The critical field is also somewhat smaller in this case $\tilde{H}_c \approx 3.8 (J/J')^{1/2}\tilde{D}$,
 mainly due to the fact that $h$ is larger by a factor of 2, see Eq.~(11).
 For other, less symmetric field directions, the form of the effective Hamiltonian,
 and consequently the field-induced patterns can be quite complex.
 Finally, in the case of a field ${\bf H} \sim (1,1,1)$, when $h=0$,
the quantum transition described above does not take place, and
the chiral  order of Fig.~\ref{Fig3}(a) essentially  does not evolve.
In this case the various additional terms similar to the ones
 described in Appendix B may lead to small, sub-leading deviations from the perfect
 chiral sate.

In addition to the field-induced ordered patterns
of Fig.~\ref{Fig3} and Fig.~\ref{Fig4}, determined mostly by the inter-tetrahedral interactions,  
 a single tetrahedron with DM interactions also  possesses a finite
 moment in the direction of the field, \cite{K} meaning that
 the spins in Fig.~\ref{Fig3} and Fig.~\ref{Fig4} would also
 tend to  tilt in that direction.
However the moment along the field is proportional to $\tilde{D}^{2}\tilde{H}$,
 as can be deduced from the fact that the ground state
 energy varies as $\tilde{D}^{2}\tilde{H}^{2}$ from (A3).
Consequently this component has not been taken into account in Eqs.~(\ref{operatorsDM},A2),
 valid to lowest order in $\tilde{D},\tilde{H}$. Finally, we emphasize that while
we have assumed $\tilde{D},\tilde{H}$ to be small, the ratio $\tilde{D}/\tilde{H}$ can
 be arbitrary, meaning that the quantum transitions in a field 
 are within the limit of  validity of our approach.

\section{Phase Diagram and Discussion}
\begin{figure}[ht]
\centering
\includegraphics[height=175pt, keepaspectratio=true]{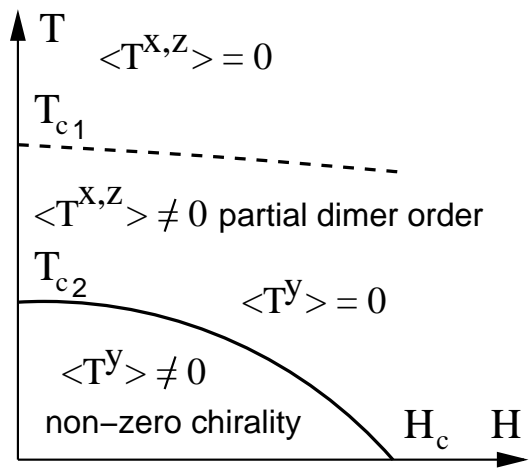}
\caption{Schematic phase diagram at non-zero temperature in the presence of small magnetic
field ($H \ll J$) and DM interactions.}
\label{Fig5}
\end{figure}

At finite temperature we expect the phase diagram to look as presented 
in Fig.~\ref{Fig5} (it is assumed that $h\neq0$). 
 The higher transition temperature $T_{c1} \sim J'^{3}/J^{2}$
corresponds to the scale below which  the translational symmetry is broken
(dimerization occurs), and is determined by the  energy scale
 in $\hat{{\cal H}}_{\mbox{eff}}$ for $D=0$, Eq.~(2). 
 We expect $T_{c1}$ to have weak dependence on magnetic field when DM interactions
 are present.
 At a lower scale $T_{c2}$ the  Ising $T^{y} \rightarrow -T^{y}$ symmetry is spontaneously broken
by Eq.~(\ref{phamDM}).
 For $H=0$ we can estimate  $T_{c2} \sim J' \tilde{D}^{2}(D/J')^{2}$. At fixed field
this finite-temperature transition is in the
 3D Ising universality class, and the specific heat diverges as 
$C\sim |T-T_{c2}(H)|^{-\alpha}, \ \alpha \approx 0.11$. \cite{GZZ}
 We emphasize that  Fig.~\ref{Fig5}  shows only the low-field part
 of the phase diagram (since $H_c \sim D \ll J$), while the physics at high fields
 cannot be determined within the effective Hamiltonian framework presented here.

\noindent

In certain pyrochlores, such as the gadolinium titanium oxides 
with S=7/2,   field-driven phase transitions have been observed, \cite{RP}
 although in this material the magnetic order is typically explained as originating
from the long-range dipolar interactions. 
 For such large value of the spin the DM mechanism for magnetic order,
at least the way  it is developed in this work, should not be effective
since our calculations were based on  strong singlet correlations in the ground state.
 At the moment it is hard to point out a class of materials
 where the DM interactions are definitely expected to be dominant with respect to 
other anisotropies capable to produce ordering; some possible
 examples are given in Ref.~\onlinecite{MCL}.
In particular, our results are specific to the case $S=1/2$, while
most currently known pyrochlores have higher $S$.
Let us also point out the main differences between the present work
 and the purely classical model: \cite{MCL} (a.) We have found that
 the antiferromagnetic order is {\it weak}, determined by the
 DM interaction scale itself, (b.) The "chiral" non-coplanar
 pattern is the stable one for $S=1/2$ in the absence of a field.
Field-induced patterns then dominate for $H > H_{c} \sim D$.
Let us also give some estimates: if we take the optimistic
 viewpoint and apply our formulas to the case
$J=J'$, and take $J \sim 100K, D/J \sim 10^{-1}$, then the
 characteristic temperature for onset of chiral order would be
$T_{c2} \sim 10mK$, and the magnitude of the moment
$|\langle {\bf S} \rangle| \sim 0.1$. The characteristic
 critical fields would be $H_{c}  \sim  10K$. These should be viewed as order
 of magnitude estimates. Inevitably the  scale $T_{c2}$ falls into the $mK$ range,
which, in combination with the smallness of  $\langle {\bf S} \rangle$ itself,
 would probably make the chiral state rather hard to observe with
 neutron scattering techniques used to probe the spin structure. \cite{Lee}
The  spectral weight of magnetic excitations in such neutron scattering measurements would be
 determined by $D$ itself. 
 However the  scale of  $H_{c}$ of around $10T$ or smaller
(also determined by $D$), means that
 the various  field-induced patterns, strongly dependent on the magnetic field
 direction, might be accessible. 

In conclusion, we have shown that DM interactions can induce weak
 antiferromagnetic order characterized by non-zero chirality. In an external
 magnetic field  quantum transitions between the chiral state and 
field-induced ordered states take place.
Field-induced patterns with different symmetries
depending on the direction of the field are very characteristic
of the presence of DM interactions. 
 We have used  an expansion around a configuration
which breaks the lattice symmetry \cite{T} and leaves a macroscopic degeneracy, subsequently
lifted by the DM interactions.
 Full restoration of lattice symmetry within such an approach seems impossible  to achieve,
 as it is impossible for example in the large-N approach. \cite{TMS1}
Nevertheless we expect that without DM interactions 
 the ground state properties  and the inherent degeneracy present in this strongly
frustrated magnet are well accounted for.  
In this situation the DM interactions ``push" the pyrochlore lattice towards
the ordered states analyzed in the present work.
 More generally,  
  the DM interactions can be relevant and lead to weak antiferromagnetism 
 in strongly frustrated systems, where the Heisenberg exchanges on their own fail to produce long-range
 order. We also emphasize that the physics behind the weakly antiferromagnetic
 states with different symmetries  discussed in this work is  very different
 from the phenomenon of weak ferromagnetism, usually associated with
 DM interactions. 
 
\begin{acknowledgments}  
Stimulating discussions with H. Tsunetsugu, O. Tchernyshyov and C. Lhuillier,   
and the financial support of the Swiss National Fund and MaNEP (V.N.K. and F.M.) are gratefully acknowledged. 
\end{acknowledgments}

\appendix
\section{Spin operators and energy levels in magnetic field}

 Here we present the expressions for the spin operators 
for magnetic field in the $z$ direction. The two ground state
 wave-functions $|s_1 \rangle, |s_1 \rangle$ are modified in the following
 way  in the presence of DM interactions
 ($|s_1 \rangle \rightarrow |\Phi \rangle$, $|s_2 \rangle \rightarrow |\Psi \rangle$):
\begin{eqnarray}
{\bf H} & = & H(0,0,1); \nonumber \\
|\Phi \rangle &  =  & |s_1 \rangle +
\frac{i \sqrt{6} \tilde{D}}{4} \left[|p_x\rangle -|p_y\rangle +|q_x\rangle
 +|q_y\rangle \right] \nonumber \\
&& + \frac{\sqrt{6} \tilde{D} \tilde{H}}{4}  \left[|p_x\rangle +|p_y\rangle -|q_x\rangle
+|q_y\rangle \right], \nonumber \\
|\Psi \rangle & = & |s_2 \rangle +
\frac{i\tilde{D}}{2\sqrt{2}}\left[|p_x\rangle +|p_y\rangle +|q_x\rangle
 -|q_y\rangle \right]  + i \tilde{D} |t_z\rangle   \nonumber\\
&& + \frac{\tilde{D} \tilde{H}}{2\sqrt{2}}  \left[-|p_x\rangle +|p_y\rangle +|q_x\rangle
+|q_y\rangle \right],
\end{eqnarray}
where $|p_{\mu}\rangle, |q_{\mu}\rangle, |t_{\mu}\rangle$, $\mu =x,y,z$ are the three excited triplet states
 on a tetrahedron. From these equations we obtain
\begin{eqnarray}
S_{1,3}^x  &= & \mp\frac{2\tilde{D}}{\sqrt{6}}T^{y} \pm \frac{\tilde{D}\tilde{H}}{\sqrt{2}}T^{z}
\pm \frac{\tilde{D}\tilde{H}}{\sqrt{6}}T^{x} \nonumber \\
S_{1,3}^y & = & \mp\frac{2\tilde{D}}{\sqrt{6}}T^{y} \pm \frac{\tilde{D}\tilde{H}}{\sqrt{2}}T^{z}
\mp \frac{\tilde{D}\tilde{H}}{\sqrt{6}}T^{x} \nonumber \\
S_{1,3}^z & = & \frac{2\tilde{D}}{\sqrt{6}}T^{y} \nonumber \\
S_{2,4}^x & = & \mp\frac{2\tilde{D}}{\sqrt{6}}T^{y} \mp \frac{\tilde{D}\tilde{H}}{\sqrt{2}}T^{z}
\mp \frac{\tilde{D}\tilde{H}}{\sqrt{6}}T^{x} \nonumber \\
 S_{2,4}^y & = & \pm\frac{2\tilde{D}}{\sqrt{6}}T^{y} \pm \frac{\tilde{D}\tilde{H}}{\sqrt{2}}T^{z}
\mp \frac{\tilde{D}\tilde{H}}{\sqrt{6}}T^{x} \nonumber \\
S_{2,4}^z & = & -\frac{2\tilde{D}}{\sqrt{6}}T^{y}.
\end{eqnarray}

We also give the ground state energy splitting on a single 
tetrahedron, in (weak)
 external magnetic field with arbitrary direction 
\begin{eqnarray}
\lefteqn{{\bf H} = (H_{x},H_{y},H_{z})} \nonumber \\
\lefteqn{E_{0}^{(1,2)}= -\frac{3J}{2} -\frac{3D^{2}}{2J} - \frac{D^{2}|{\bf H}|^{2}}{J^{3}}} \nonumber \\
&&   \pm \frac{D^{2}}{2J^{3}}  \sqrt{ |{\bf H}|^{4} -
 3(H_{x}^{2}H_{y}^{2}+H_{x}^{2}H_{z}^{2}+H_{z}^{2}H_{y}^{2})}  \nonumber \\ 
\end{eqnarray}
From here the ``effective magnetic field"  $h$ appearing in the
 pseudospin Hamiltonian (10) is: $h =  E_{0}^{(2)} - E_{0}^{(1)}$. 

\section{Effective Hamiltonian contributions in magnetic field}
 
 We briefly discuss the structure and treatment of the term $\delta \hat{{\cal H}}_{\mbox{eff}}^{(H)}$
in (10). As is clear from (7) and (A2), potentially contributions of two types appear:
 (a.) terms of order $\tilde{D}^{2} \tilde{H}$, and (b.) terms of order  $\tilde{D}^{2} \tilde{H}^{2}$.
 While treating these terms we will assume that the unit cell structure  
 of  Fig.~2(a) does not change. Indeed, since we are interested in 
 the case of weak  fields $\tilde{H} \ll 1$, the above coupling
 constants are of sub-leading order with respect to the case without a field, and therefore
 one does not expect the unit cell to change. Under this assumption we find that the
 terms of order $\tilde{D}^{2} \tilde{H}$ vanish identically for the field directions
 considered in this work.   The remaining contribution, e.g.  
 for a field ${\bf H}= \frac{H}{\sqrt{2}}(1,1,0)$, written (per site) within
 the unit cell of  Fig.~2(a) is  
\begin{equation}
\langle \delta \hat{{\cal H}}_{\mbox{eff}}^{(H)} \rangle = \frac{J'}{6} \tilde{D}^{2} \tilde{H}^{2} \ 
\delta \hat{{\cal H}} ,  
\end{equation}
where
\begin{eqnarray}
\delta \hat{{\cal H}} & = &  - 4(T_{0}^{x}+T_{3}^{x})(T_{1}^{x}  + T_{2}^{x}) +  \nonumber \\
&&\sqrt{3}(T_{0}^{x}-T_{3}^{x})(T_{1}^{z}  -  T_{2}^{z}) +\nonumber \\
&& \sqrt{3}  (T_{0}^{z}-T_{3}^{z}) (T_{1}^{x} -  T_{2}^{x}) + \nonumber \\
&& 3(T_{0}^{x}T_{3}^{x} - T_{0}^{z} T_{3}^{z} 
 +T_{1}^{x}T_{2}^{x} -T_{1}^{z}T_{2}^{z}).
\end{eqnarray}
These  terms were taken in to account in our numerical solution of the mean-field
 equations  corresponding to (10).


\begin{thebibliography}{99}
\bibitem{Remark} Our choice of ground states is different from that
 of Ref.~\onlinecite{T} leading to  different coefficients in $\hat{{\cal H}}_{\mbox{eff}}$.
\end{thebibliography}
\end{document}